\documentstyle[aps,prl,twocolumn,epsf,rotate]{revtex}
\begin{document}
\flushbottom
\draft
\twocolumn[\hsize\textwidth\columnwidth\hsize\csname
@twocolumnfalse\endcsname
\title{Long-range order and low-energy spectrum of diluted 2D quantum AF}
\author{A. L. Chernyshev\cite{perm}, Y. C. Chen,
and A. H. Castro Neto}
\address{
Department of Physics, University of California, Riverside, CA 92521}
\date{\today}
\maketitle

\widetext\leftskip=1.9cm\rightskip=1.9cm\nointerlineskip\small
\begin{abstract}
\hspace*{2mm}
The problem of diluted two-dimensional (2D) quantum antiferromagnet
(AF) on a square lattice is studied using spin-wave theory. 
The influence of impurities on static and dynamic properties 
is investigated and a good agreement with experiments and
Monte Carlo (MC) data is found. 
The hydrodynamic description of spin waves breaks down at
characteristic wavelengths $\Lambda\agt\exp(\frac{const}{x})$, $x$ being
an impurity concentration, while
the order parameter is free from anomalies. 
We argue that this dichotomy originates from strong 
scattering of the low-energy excitations in 2D. 
\end{abstract}
]
\narrowtext
The interest in magnetic properties of high-T$_c$ compounds
has been a major driving force of intensive studies of
low-dimensional magnetic systems during the last decade \cite{CHN}. 
One of such systems is the La$_2$Cu$_{1-x}$Zn(Mg)$_x$O$_4$, a 
parental compound doped with {\it static} vacancies, which  
shows greater stability of the AF order against doping  
than its mobile hole doped counterpart La$_{2-x}$Sr$_x$CuO$_4$ 
\cite{early_exp,Greven}. 
It represents a fine example of a diluted 2D quantum $S=1/2$
AF whose properties are the subject of this study.
The problem of diluted spin systems has attracted much
attention in the past \cite{Harris,Shender} and their physics is well
understood. The general understanding 
is that the low-energy excitations of these systems are weakly damped
spin waves which belong to the infinite cluster \cite{Shender} and are
well defined up to the percolation threshold.
In this work we show that quantum effects and low dimensionality
prevent the excitation spectrum to be defined in these hydrodynamic
terms \cite{HH} at arbitrarily small doping
and lead to a paradoxical situation where the long-range order is
preserved but the long-wavelength spectrum is not ballistic.

The site-diluted AF on a square lattice is described by:
\begin{eqnarray}
\label{H}
{\cal H} = J\sum_{\langle ij \rangle} p_i p_j {\bf S}_i\cdot {\bf S}_j\ ,
\end{eqnarray}
with $p_i=1 (0)$ for the magnetic (non-magnetic) site. 
The pure system at long wavelengths can be described by the
nonlinear $\sigma$-model \cite{CHN}.
Its applicability in the presence of a quenched disorder has been
questioned \cite{CHN} since impurities destroy the Lorentz
invariance of the model.  
Some generalizations of the $\sigma$-model  
have nevertheless been proposed  with parameters
modified according to MC data \cite{Man1} and the
classical percolation theory \cite{Chen}. 

In this work we study the problem of disorder using the 
$t$-matrix approach combined with a configurational average 
over the random 
positions of impurities. It also leads to description of
the system in terms of an effective medium with renormalized
parameters. In an earlier work Harris and Kirkpatrik \cite{Harris}
have shown that in 2D a non-hydrodynamic term appears in the spin-wave
self-energy which explicitly violates Lorentz invariance: 
$\Sigma_{\bf k}(\omega)\propto x k (\ln|\omega|-i\pi/2)$. A recent
study by Wan {\it et al}. \cite{Wan} has confirmed this result.
One can show that the spectrum is overdamped at wavelengths
$\Lambda\agt\exp(\pi/4x)$ which implies the existence of a
new lengthscale in the system. The absence of a well-defined 
long-wavelength mode also suggests that the true order should be
unstable under an infinitesimal impurity doping, or that  $d=2$ is the
upper critical dimension for this problem\cite{Harris}.

In this work we study the effect of impurities on the spectrum, staggered
magnetization, $M(x)$, and N\'{e}el temperature,
$T_N(x)$, to clarify the problem of
stability of a (quasi-)2D AF order. We show that neither $M(x)$ nor $T_N(x)$
possesses anomalous contributions which would imply an instability.
Thus, we have a somewhat paradoxical situation
when the spectrum is ill-defined
while the order parameter is not affected.  
Such a dichotomy comes from the strong influence of disorder in the
low-energy excitations in 2D. The
averaging procedure, which effectively restores translational
invariance, fails to recover the long-wavelength excitation
spectrum of this effective medium \cite{d_wave}. The low-frequency 
modes do exist in some form but they
cannot be classified in terms of an effective
wave-vector and thus the long-wavelength propagation
is entirely diffusive. This scheme implies an
existence of a disorder-induced energy scale which restricts the
applicability of the continuum approach.

We start with the Hamiltonian (\ref{H}) in the spin-wave
approximation ${\cal H}={\cal H}_0+{\cal H}_{imp}^A+
{\cal H}_{imp}^B$:
\begin{eqnarray}
\label{H_0}
&&{\cal H}_0 \simeq 4SJ\sum_{\bf k}\omega_{\bf k}
(\alpha^{\dag}_{\bf k}\alpha_{\bf k}+\beta^{\dag}_{\bf k}\beta_{\bf
k})\ ,\\
\label{H_imp}
&&{\cal H}_{imp}^A \simeq -4SJ\sum_{l\in A,{\bf k},{\bf k}^\prime}
e^{i({\bf k}-{\bf k}^\prime){\bf R}_l}\bigl[
V^{\alpha\alpha}_{A,{\bf k k}^\prime}~\alpha^{\dag}_{\bf k}
\alpha_{{\bf k}^\prime}\\
&&\phantom{{\cal H}_{imp}^A \simeq -4S}
+V^{\beta\beta}_{A,{\bf k k}^\prime}~\beta^{\dag}_{\bf k}\beta_{{\bf
k}^\prime}
+V^{\alpha\beta}_{A,{\bf k k}^\prime}~(\alpha^{\dag}_{\bf k}
\beta^{\dag}_{-{\bf k}^\prime}+\mbox{H.c.})
\bigr]
\ ,\nonumber
\end{eqnarray}
where the quadratic part of the pure host Hamiltonian ${\cal H}_0$ is
diagonalized using Bogolyubov transformation:
$S^{\dag}_{{\bf k},A}/\sqrt{2S}\simeq a_{\bf k}=u_{\bf k}
\alpha_{\bf k}+v_{\bf k}\beta^{\dag}_{-{\bf k}}$,
$S^{\dag}_{{\bf k},B}/\sqrt{2S}\simeq b^{\dag}_{\bf k}=u_{\bf k}
\beta^{\dag}_{\bf k}+v_{\bf k}\alpha_{-{\bf k}}$, 
$A$ and $B$ denote the sublattices, with $u_{\bf
k}^2-v_{\bf k}^2=1$, $2u_{\bf k}v_{\bf k}=-\gamma_{\bf k}/\omega_{\bf
k}$, $\gamma_{\bf k}=(\cos(k_x)+\cos(k_y))/2$, 
and bare spin-wave frequency $\omega_{\bf k}=\sqrt{1-\gamma_{\bf k}^2}$. 
All momenta belong to the magnetic Brillouin zone $k_x+k_y\le\pi$.
${\cal H}_{imp}^B={\cal H}_{imp}^A(A\rightarrow B)$, interactions are
given by 
$V^{\alpha\alpha}_{A}=V^{\beta\beta}_{B}=u_1u_2+\gamma_{1-2}v_1v_2
+\gamma_1v_1u_2+\gamma_2v_2u_1$, $V^{\beta\beta}_{A}=
V^{\alpha\alpha}_{B}=
v_1v_2+\gamma_{1-2}u_1u_2+\gamma_1u_1v_2+\gamma_2u_2v_1$, and 
$V^{\alpha\beta}_{A,1,2}=V^{\alpha\beta}_{B,2,1}=u_1v_2+
\gamma_{1-2}v_1u_2+\gamma_1v_1v_2+\gamma_2u_2u_1$, $l$ runs
over the impurity sites. 
As a first step we solve the scattering problem for the single
impurity. Since the impurity potential is short-ranged 
one uses the $t$-matrix approach.
The advantage of this method
is that the single impurity problem can be
solved {\it exactly} (within the spin-wave approximation)
\cite{Izyumov}. Therefore, the results of this approach become
exact in the dilute limit $x\rightarrow 0$.

In our case the square lattice site-defect scatters only in $s$-,
$p_x$-, $p_y$-, and $d$-symmetric channels and one needs to solve
the $t$-matrix for each harmonic. 
The $t$-matrix equations are:
\begin{eqnarray}
\label{t_matrix}
&&\Gamma_{A,i}^{\alpha\alpha}=-V_{A,i}^{\alpha\alpha}-
V_{A,i}^{\alpha\alpha}G^0_{\alpha}\Gamma_{A,i}^{\alpha\alpha}-
V_{A,i}^{\alpha\beta}G^0_{\beta}
\Gamma_{A,i}^{\alpha\beta}\ ,\nonumber\\
&&\Gamma_{A,i}^{\alpha\beta}=-V_{A,i}^{\alpha\beta}-V_{A,i}^{\beta\beta}
G^0_{\beta}\Gamma_{A,i}^{\alpha\beta}-V_{A,i}^{\alpha\alpha}G^0_{\alpha}
\Gamma_{A,i}^{\alpha\alpha}\ ,
\end{eqnarray}
where $V_i$'s are the $i=s, p_x, p_y, d$ components of V's
(\ref{H_imp}). We suppress ${\bf k}$ and $\omega$ in vertex
functions $\Gamma_i({\bf k},{\bf k}^\prime,\omega)$, 
bare Green's functions $G^0_{\alpha}({\bf k},\omega)=
G^0_{\beta}({\bf k},-\omega)=\{\omega-\omega_{\bf k}+i0\}^{-1}$, and
interactions $V_{{\bf kk}^\prime}$. Products $VG\Gamma$ involve
summation over the internal momenta, all energies are in units of
$4SJ$. An equivalent set of equations gives the vertices
$\Gamma_{A,i}^{\beta\beta}$. For the impurity in $B$ sublattice
$\Gamma_{B,i}\equiv\Gamma_{A,i}(\alpha\leftrightarrow\beta)$.
Since the impurity potential in each partial wave is separable:
$V_{A,s}^{\alpha\alpha}=u_{\bf k}\omega_{\bf k}u_{{\bf k}^\prime}
\omega_{{\bf k}^\prime}$,
$V_{A,p_\sigma}^{\alpha\alpha}=\phi_{k_\sigma}v_{\bf k}
\phi_{k^\prime_\sigma}v_{{\bf k}^\prime}/2$,
$V_{A,d}^{\alpha\alpha}=\gamma^-_{\bf k}v_{\bf k}
\gamma^-_{{\bf k}^\prime}v_{{\bf k}^\prime}$,
$V_{A,i}^{\beta\beta}=V_{A,i}^{\alpha\alpha}(u\leftrightarrow v)$,
$V_{A,s}^{\alpha\beta}=-u_{\bf k}\omega_{\bf k}v_{{\bf k}^\prime}
\omega_{{\bf k}^\prime}$,
$V_{A,p_\sigma}^{\alpha\beta}=\phi_{k_\sigma}v_{\bf k}
\phi_{k^\prime_\sigma}u_{{\bf k}^\prime}/2$,
$V_{A,d}^{\alpha\beta}=\gamma^-_{\bf k}v_{\bf k}
\gamma^-_{{\bf k}^\prime}u_{{\bf k}^\prime}$, where
$\phi_{k_\sigma}=\sin(k_\sigma)$, 
$\gamma^-_{\bf k}=(\cos(k_x)-\cos(k_y))/2$,
Eqs. (\ref{t_matrix}) can be readily solved: 
$\Gamma_{A,i}^{\sigma\sigma^\prime}({\bf k},{\bf k}^\prime,\omega)
=-V_{A,i}^{\sigma\sigma^\prime}({\bf k},{\bf k}^\prime)
\tilde{\Gamma}_{i}(\omega)$, 
where
\begin{eqnarray}
\label{gammas}
&&\tilde{\Gamma}_{s}(\omega)=-1/\omega
-(1+\omega)\rho(\omega)/[1-\omega(1+\omega)\rho(\omega)]\ ,\nonumber\\
&&\tilde{\Gamma}_{p}(\omega)=2/
[1+\omega+(1-\omega)(\omega^2\rho(\omega)-\rho_d(\omega))]\ ,\\
&&\tilde{\Gamma}_{d}(\omega)=1/[1+(1-\omega)\rho_d(\omega)]
\ ,\nonumber
\end{eqnarray}
with $\rho=\sum_{\bf p}1/(\omega^2-\omega_{\bf p}^2)$ and 
$\rho_d=\sum_{\bf p}(\gamma^-_{\bf p})^2/(\omega^2-\omega_{\bf p}^2)$, 
which can be expressed through the complete elliptic integrals
\cite{Chen_future}. For the impurity in $B$ sublattice
$\Gamma_{B,i}^{\sigma\sigma^\prime}=-V_{B,i}^{\sigma\sigma^\prime}
\tilde{\Gamma}_{i}(-\omega)$.
The $s$-wave scattering (\ref{gammas}) reveals a zero-frequency mode 
which originates from the oscillations of the  
fictitious degrees of freedom at the impurity site. 
Roughly speaking, since in the spin-wave approximation spins are 
quantized through bosons even if $S^\prime$ at the impurity site 
is set to zero there is still $a^\dag_0 a_0$ left in $S^z_0$ (for
discussion see Ref. \cite{Wan}).    
This gives rise to the unphysical
zero-frequency mode which has to be projected out. 
We do so by introducing magnetic fields at the impurity 
sites (similar to Refs. \cite{BK1,Leonid})
$\Delta{\cal H}=H_z\sum_l a^{\dag}_i a_i$. Within this approach 
after straightforward algebra \cite{Chen_future} in the
limit $H_z\rightarrow\infty$ one obtains: 
$\Gamma_{A,s}^{\sigma\sigma^\prime}({\bf k},{\bf k}^\prime,\omega)
=-V_{A,s}^{\sigma\sigma^\prime}({\bf k},{\bf k}^\prime)
\tilde{\Gamma}_{s}(\omega)+
\Delta\Gamma_{A,s}^{\sigma\sigma^\prime}({\bf k},
{\bf k}^\prime,\omega)$, 
\begin{eqnarray}
\label{gamma_s}
\tilde{\Gamma}_{s}(\omega)=
-(1+\omega)\rho(\omega)/[1-\omega(1+\omega)\rho(\omega)]\ ,
\end{eqnarray}
which is free from the zero-frequency pole. $\Delta\Gamma_{A,s}^{\alpha\alpha}
=u_{\bf k}u_{{\bf k}^\prime}[\omega_{\bf k}+\omega_{{\bf k}^\prime}
-\omega]$, $\Delta\Gamma_{A,s}^{\beta\beta}
=-v_{\bf k}v_{{\bf k}^\prime}[\omega_{\bf k}+\omega_{{\bf k}^\prime}
+\omega]$, $\Delta\Gamma_{A,s}^{\alpha\beta}
=-u_{\bf k}v_{{\bf k}^\prime}[\omega_{\bf k}-\omega_{{\bf k}^\prime}
+\omega]$.
$\Gamma_{B}=\Gamma_A(-\omega)\{\alpha\leftrightarrow\beta\}$ as
before, $p$ and $d$ waves are not affected by the projection.

The next step is the averaging procedure which restores
translational invariance. Assuming random distribution of impurities
one readily transforms scattering vertices into the spin-wave
self-energies: $\Sigma_{\bf k}^{\sigma\sigma^\prime}(\omega)=
\sum_i\Sigma_{i,{\bf k}}^{\sigma\sigma^\prime}(\omega)$, with 
$\Sigma_{i,{\bf k}}^{\sigma\sigma^\prime}(\omega)=
x\delta_{{\bf k}-{\bf k}^\prime}[\Gamma_{A,i}^{\sigma\sigma^\prime}
({\bf k},{\bf k}^\prime,\omega)+
\Gamma_{B,i}^{\sigma\sigma^\prime}({\bf k},{\bf k}^\prime,\omega)]$
\begin{eqnarray}
\label{sigmas}
&&\Sigma_{s,{\bf k}}^{\alpha\alpha}(\omega)/x=-\omega_{\bf k}^2\bigl[
u_{\bf k}^2\tilde{\Gamma}_{s}(\omega)+
v_{\bf k}^2\tilde{\Gamma}_{s}(-\omega)\bigr]
+2\omega_{\bf k} -\omega\ ,\nonumber\\
&&\Sigma_{p(d),{\bf k}}^{\alpha\alpha}(\omega)=xA^{p(d)}_{\bf k}
\bigl[v_{\bf k}^2\tilde{\Gamma}_{p(d)}(\omega)+
u_{\bf k}^2\tilde{\Gamma}_{p(d)}(-\omega)\bigr]\ ,\\
&&\Sigma_{i,{\bf k}}^{\alpha\beta}(\omega)=xB^i_{\bf k}
\bigl[\tilde{\Gamma}_{i}(\omega)+
\tilde{\Gamma}_{i}(-\omega)\bigr]\ ,\
\Sigma_{i,{\bf k}}^{\beta\beta}(\omega)=
\Sigma_{i,{\bf k}}^{\alpha\alpha}(-\omega) ,
\nonumber
\end{eqnarray}
where $A^p_{\bf k}=-\omega_{\bf k}^2+(\gamma_{\bf k}^-)^2$,
$A^d_{\bf k}=-(\gamma_{\bf k}^-)^2$, $B^s_{\bf k}=\omega_{\bf k}^2
u_{\bf k}v_{\bf k}$, $B^p_{\bf k}=A^p_{\bf k}u_{\bf k}v_{\bf k}$, 
$B^d_{\bf k}=A^d_{\bf k}u_{\bf k}v_{\bf k}$. It is interesting that 
``on shell'' ($\omega=\omega_{\bf k}$) 
Eqs. (\ref{gammas}) and (\ref{gamma_s}) yield identical
$\Sigma_{s,{\bf k}}(\omega_{\bf k})$. 

\begin{figure}
\unitlength 1cm
\epsfxsize=8.cm
\begin{picture}(8,2.4)
\put(0.5,0.3){\epsffile{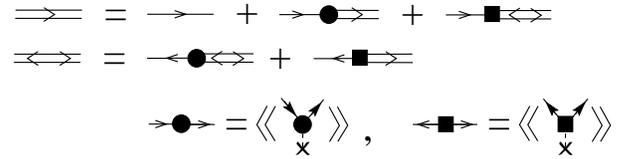}}
\end{picture}
\caption{Belyaev diagram series for the diagonal, $G^{\alpha\alpha}$, and
off-diagonal,  $G^{\alpha\beta}$, Green's functions. Self-energies 
$\Sigma^{\alpha\alpha}$ (circle) and $\Sigma^{\alpha\beta}$
(square) are the configurational averages of $\Gamma^{\alpha\alpha}$
and $\Gamma^{\alpha\beta}$, respectively.}
\label{fig1}
\end{figure}

At low energies $\omega, \omega_{\bf k}\rightarrow 0$ self-energies
are given by:
\begin{eqnarray}
\label{sigma_le}
&&\Sigma_{\bf k}^{\alpha\alpha}(\omega)\simeq x\omega_{\bf k}\bigl[
\rho(\omega)+2-\pi/2\bigr]-x\omega+
{\cal O}(\omega_{\bf k}\omega^2\rho^3)\ ,
\nonumber\\
&&\Sigma_{\bf k}^{\alpha\beta}(\omega)\simeq x\omega_{\bf k}\bigl[
\rho(\omega)+\pi/2\bigr]+{\cal O}(\omega_{\bf k}\omega^2\rho^3)\ , \\
&&\mbox{with}\ \ \ \rho(\omega)\simeq (2/\pi)\ln(|\omega|/4)-i\ ,
\nonumber
\end{eqnarray}
which includes contributions from $s$- and $p$-wave scattering,
$\Sigma_d\simeq {\cal O}(\omega_{\bf k}^3)$. One can see that besides
the ``normal'' softening of the long-wavelength mode and the damping
proportional to the higher power of $k$ ($\sim k^3$) one acquires a
non-linear dispersion term with the damping $\tilde{\gamma}_{\bf
k}/\omega_{\bf k}\sim x$ having only parametric
smallness with respect to the bare spectrum. 
Renormalization of the real part of the spectrum is dominated by the
$\ln(|\omega|)$ term at low frequencies.
A naive ``on-shell'' pole equation
$\tilde{\omega}^{\prime}_{\bf k}=\omega_{\bf k}+\Re\Sigma_{\bf
k}^{\alpha\alpha}(\omega_{\bf k})$
suggests a vanishing of the spectrum \cite{Wan} and an
instability of the ground state. Consideration of the 
diagrammatic series for the Green's functions shown in
Fig. \ref{fig1} with self-energies defined in Eq. (\ref{sigmas}) and
$G^{11}_{\bf k}(\omega)=\{
\omega-\omega_{\bf k}-\Sigma_{\bf k}^{\alpha\alpha}(\omega)\}^{-1}$
gives:
\begin{eqnarray}
\label{G_full}
&&G_{\bf k}^{\alpha\alpha}(\omega)=\frac{G^{11}_{\bf k}(-\omega)^{-1}}
{[G^{11}_{\bf k}(\omega)G^{11}_{\bf k}(-\omega)]^{-1}-
(\Sigma_{\bf k}^{\alpha\beta}(\omega))^2}\ ,
\nonumber\\
&&G_{\bf k}^{\alpha\beta}(\omega)=
\frac{\Sigma_{\bf k}^{\alpha\beta}(\omega)}
{[G^{11}_{\bf k}(\omega)G^{11}_{\bf k}(-\omega)]^{-1}-
(\Sigma_{\bf k}^{\alpha\beta}(\omega))^2}\ .
\end{eqnarray}
At $k\gg\omega_0\sim\exp(-\pi/4x)$ spectral function reveals 
damped quasiparticle peak at $\omega\alt k$
plus some structure at low frequencies $\omega\sim\omega_0$. At
low $k\sim \omega_0$ the quasiparticle peak merges with this structure
and disappears. The spectrum is completely overdamped 
at $k\alt\omega_0$. 
Beyond this scale no hydrodynamic description is possible.

A potential concern is whether an anomaly in Eq. (\ref{sigma_le})
can be fixed by the higher order terms.
General discussion of this problem in terms of the
susceptibility (Ref.\cite{Harris}) concluded that the divergency
should be unrelated to the expansion in $x$. One
can also show that the summation of the Green's
function series (\ref{G_full}) and dressing of the inner lines
in self-energies using self-consistent equation on $\rho(\omega)$ and
$G^{\sigma\sigma}$  actually enhance the anomaly.

In a complementary problem of a quasi-2D AF a small inter-plane
coupling $\alpha=J_\perp/J$ cuts off the log-singularity:
$\rho_{3D}(\omega)\simeq (1/\pi)\ln(\alpha/32)+i{\cal O}(\omega)$ 
at $\omega<\sqrt{2\alpha}$. Therefore a ``safe'' range of concentrations
$x<x^*\sim\ln^{-1}(1/\alpha)$ can be found where the long-wavelength
quasiparticles are still well defined deep in the 3D region of the 
${\bf k}$-space ($\Lambda^{-1}\ll\sqrt{\alpha}$). 
However, one should be able to observe a
nonlinearity of the spectrum and an abnormal damping of the spin waves
in the 2D long-wavelength region ($\sqrt{\alpha}<\Lambda^{-1}\ll 1$) 
similar to the quasi-1D problem \cite{SK}. For the
real materials $\alpha\sim 10^{-4}$ giving $x^*\sim 20\%$.
Above the concentration $x^*$ all the low-energy excitations are
incoherent because the 2D disorder-induced scale $\omega_0^{-1}$ is 
shorter than the 3D length $ 1/\sqrt{\alpha}$ so
the spin waves lose coherence 
before they can propagate in 3D. The same consideration
applies to the case of small anisotropies introducing gaps in the spectrum 
with a modified $\alpha=\alpha_{eff}$ accumulating the total effect 
of the gaps and 3D coupling.
It should be noted that the incoherence comes
from the averaging procedure which converts the dissipation
of momentum into the dissipation of the energy. Therefore, the overdamped
excitations should be understood as diffusive.
It is interesting that it requires 2D and ``strong'' disorder to restrict
the number of Euclidean paths for spin waves and 
to break down the description of the problem in terms of an effective
medium. 

\begin{figure}
\unitlength 1cm
\epsfxsize=6.cm
\begin{picture}(7,6)
\put(-0.1,0.8){\rotate[r]{\epsffile{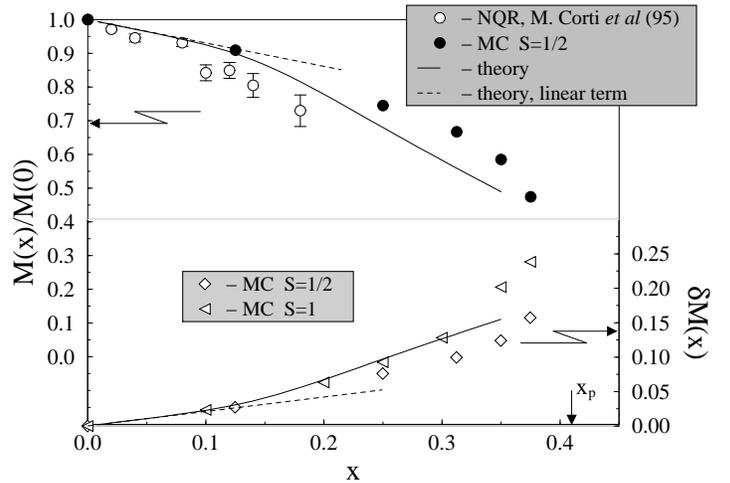}}}
\end{picture}
\caption{Left axes: reduction of the local magnetic moment by the quantum 
fluctuations induced by impurities $M(x)/M(0)$ from NMR ($\circ$,
Ref.\protect\cite{Corti}) and MC ($\bullet$, Ref.\protect\cite{Kato})
studies. Lines show our results from Eq. (\ref{M}). Right axes:
the absolute value of $\delta M(x)$ from Eq. (\ref{M}) (lines) and 
MC data, Ref.\protect\cite{Kato}, for $S=1/2$ ($\diamond$) and $S=1$ 
($\triangleleft$).}
\label{fig2}
\end{figure}

Now we proceed with quantities whose expansion in $x$ can be shown to
be reliable. Staggered magnetization at the magnetic site 
$M(x)=M_0-\delta M(x)$, $M_0=S-\delta\lambda$,
\begin{eqnarray}
\label{M}
&&\delta M(x)=\sum_{\bf k}
[\langle\alpha^\dag_{\bf k}\alpha_{\bf k}\rangle-\gamma_{\bf k}
\langle\alpha^\dag_{\bf k}\beta^\dag_{\bf k}\rangle]/\omega_{\bf k}=\\
&&\phantom{\delta M(x)=}-\sum_{\bf k}
\int_{-\infty}^{\infty}\frac{n(\omega)\mbox{d}\omega}{\pi\omega_{\bf k}}
\mbox{Im}[
G^{\alpha\alpha}_{R,{\bf k}}(\omega)-\gamma_{\bf k}
G^{\alpha\beta}_{R,{\bf k}}(\omega)]\ ,
\nonumber
\end{eqnarray}
where $\delta\lambda=\sum_{\bf k}v_{\bf k}^2\simeq 0.1966$ is
from zero-point fluctuations, $n(\omega)=[e^{\omega/T}-1]^{-1}$ is the
Bose occupation number, index $R$ denotes retarded. At $T=0$ 
$n(\omega)=-\theta(-\omega)$ and integrals in (\ref{M}) with 
$G^{\sigma\sigma^\prime}$ from (\ref{G_full}) can be taken numerically.
It can be shown \cite{Chen_future} that the expansion in $x$ for 
$G^{\sigma\sigma^\prime}$
can be performed before the integration
$G^{\alpha\alpha}\simeq G^0_{\alpha}+G^0_{\alpha}\Sigma^{\alpha\alpha}
G^0_{\alpha}$, $G^{\alpha\beta}\simeq 
G^0_{\beta}\Sigma^{\alpha\alpha}G^0_{\alpha}$
and that all integrals for the linear in $x$ term are convergent giving
$\delta M(x)=0.209(8)\cdot x+{\cal O}(x^2)$. For $S=1/2$ 
$M(x)/M_0\simeq 1- 0.691(5)\cdot x$. This linear slope
together with the results  of calculations using
 $G^{\sigma\sigma^\prime}$ from Eq.
(\ref{G_full}), MC \cite{Kato} and NMR data \cite{Corti} are
shown in Fig. \ref{fig2}. Note that Fig. \ref{fig2} and Eq. (\ref{M})
are showing the reduction of the magnetic moment by the quantum
fluctuations induced by impurities. In order to extract the same
quantity from the MC
data \cite{Kato}, which are averaged over all sites, one needs to 
deduct a probability of finding a spin-occupied site from them. One
can see a very good agreement of our results with numerical data up
to high concentrations. The oxidation of the crystals can be the reason
of a faster decrease of $M(x)$ in NMR data. 
The absolute value of $\delta M(x)$ is independent of $S$ in the linear
spin-wave approximation similar to the quantum reduction of $S$ by
$\delta\lambda$. We plot $\delta M(x)$ in Fig.\ref{fig2} (right axes)
in order to emphasize the agreement with the MC data for $S=1/2$ and
$S=1$, which show only weak $S$-dependence. 

\begin{figure}
\unitlength 1cm
\epsfxsize=6.cm
\begin{picture}(7,6)
\put(0.2,0.8){\rotate[r]{\epsffile{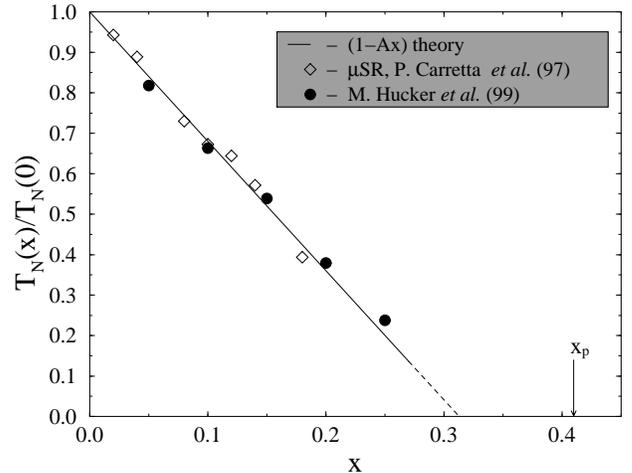}}}
\end{picture}
\caption{N\'{e}el temperature v.s. $x$ from $\mu$SR and susceptibility
measurements Refs.\protect\cite{Hucker,Carretta} for the
La$_2$Cu$_{1-x}$Zn(Mg)$_x$O$_4$, and the linear slope from our theory.}
\label{fig3}
\end{figure}

The quantity which one would expect to exhibit a different behavior is
$T_{N}$ for quasi-2D problem. The thermal corrections to the staggered
magnetization $\delta M^T\sim T\ln(T/\omega)|_{\omega=\sqrt{\alpha}}$ 
yield a finite $T_{N}\sim\ln(1/\alpha)^{-1}$ for a pure system.
Since the original spectrum is bent from the linear form this should
manifest itself in the finite temperature part of 
Eq. (\ref{M}) as a stronger divergency. One indeed finds such
contributions in Eq.(\ref{M}) ($\sim xT\ln^2|\omega|$)
already in a perturbative limit ($x\ln|\omega|\ll 1$). However,
these anomalous terms from diagonal and off-diagonal parts
cancel each other. From the mean-field equation
$M^{T_N}(x)=0$ after some algebra one obtains (at 
$\alpha\rightarrow 0$): $T_N(x)/T_N(0)\simeq 1-xA$ with $A=\pi-2/\pi+
0.209(8)/M_0$. For $S=1/2$,
$A=3.196(5)$ \cite{McGurn}, 
which fits very well experimental data \cite{Hucker,Carretta}, 
see Fig.\ref{fig3}.

Thus the order is preserved up to a high $x$ value \cite{Kato,Greven} 
while the long-wavelength spectrum is not well defined at any $x$. 
Experimentally one would need to probe directly the spin-wave spectrum
close to the AF ordering vector and look for anomalous broadening and
nonlinearity. The actual overdamped part of the spectrum is associated
with a small energy scale (and small momenta, e.g. at $x=0.2$, $k_0\sim
0.01\pi/a$) and it is not clear if it is observable
within the range of doping where conclusions of this work are
reliable. The observed deviation from the simple exponential behavior
of the correlation length $\xi(T,x)$ v.s. $1/T$ \cite{Greven} might be
related to the character of the low-energy spectrum we discuss. 

Another aspect relates our problem to the other problems
of disorder in 2D.
 We observe that some weight from every mode is
transferred to the low energies $\omega\sim\omega_0$ which can be 
understood as some sort of localization, though the localization
criteria for this problem is unclear.
 Altogether these low-energy states result in a peak in
the density of states. From $x$-expanded form of the Green's function 
$G^{\alpha\alpha}\simeq G^0_{\alpha}+G^0_{\alpha}\Sigma^{\alpha\alpha}
G^0_{\alpha}$ one readily gets $N(\omega)=2\omega/\pi +xC+{\cal
O}(x\omega\ln|\omega|)$, which implies a finite 
density of states at $\omega =0$. With $G^{\alpha\alpha}$ from
Eq. (\ref{G_full}) one obtains 
$N(\omega)\approx 2\omega/\pi+xC/[(1+4x\ln|\omega|/\pi)^2+4x^2]$
which is zero strictly at $\omega=0$ but has a sharp peak $\sim 1/x$ at 
$\omega=\omega_0$.
This is very close to the dispute over $N(\omega)$ 
for the certain types of disorder in 2D systems with 
linear excitation spectrum \cite{d_wave}. Note that in our case
there is no ``weak'' limit for disorder since the perturbation ($\sim
JSz$) is of the same strength as the pure coupling \cite{weak}. 

In conclusion, we have provided evidence for the stability of the
long-range order in the 2D and quasi-2D AF doped with static vacancies. 
At the same time the long-wavelength excitation spectrum is shown to be
diffusive for any value of doping, restricting a hydrodynamic
approach to the problem from low energies.

We would like to acknowledge enlightening discussions with 
A.~Abanov, Ar.~Abanov, A.~Balatsky, A.~Chubukov, C.~Chamon,
M.~Greven, A.~B.~Harris, O.~Vajk, L.~Pryadko, A.~Sandvik, O.~Starykh,
O.~Sushkov, and M.~Zhitomirsky. 
We are indebted to M.~Greven and O.~Vajk for sharing their results
prior to publication and to S. Todo for providing us with the MC data.
The support is provided in part by a CULAR research grant under the
auspices of the US Department of Energy.

\noindent
\noindent
\noindent


\begin{references}
\bibitem{weak} A generalization of the problem with defects altering bond 
strength gives a ``weak'' disorder when $J-J^\prime\ll J$.
This case is shown \cite{Wan} to belong to a different class and to
allow for a hydrodynamic description.
\end{references}
\end{document}